\documentclass[12pt,preprint]{aastex} 

\shorttitle{ Coronal Loop Inventory}
\shortauthors{ J.T. Schmelz et al.}

\begin{document}

\title{The Coronal Loop Inventory Project: Expanded Analysis and Results}

\author{
J.T.\ Schmelz\altaffilmark{1,2,3}, 
G.M. Christian\altaffilmark{3}, 
R.A. Chastain\altaffilmark{3}}
\altaffiltext{1}{USRA, 7178 Columbia Gateway Drive, Columbia, MD 21046, jschmelz@usra.edu}
\altaffiltext{2}{Arecibo Observatory, HC-3 Box 53995, Arecibo PR 00612}
\altaffiltext{3}{Physics Department, University of Memphis, Memphis, TN 38152}



\begin{abstract}
We have expanded upon earlier work that investigates the relative importance of coronal loops with isothermal $versus$ multithermal cross-field temperature distributions. These results are important for determining if loops have substructure in the form of unresolved magnetic strands. We have increased the number of loops targeted for temperature analysis from 19 to 207 with the addition of 188 new loops from multiple regions. We selected all loop segments visible in the 171-\AA\ images of the Atmospheric Imaging Assembly (AIA) that had a clean background. 86 of the new loops were rejected because they could not be reliably separated from the background in other AIA filters. 61 loops required multithermal models to reproduce the observations. 28 loops were effectively isothermal, that is, the plasma emission to which AIA is sensitive could not be distinguished from isothermal emission, within uncertainties. 10 loops were isothermal. Also part of our inventory were one small flaring loop, one very cool loop whose temperature distribution could not be constrained by the AIA data, and one loop with inconclusive results. Our survey can confirm an unexpected result from the pilot study: we found no isothermal loop segments where we could properly use the 171-to-193 ratio method, which would be similar to the analysis done for many loops observed with TRACE and EIT. We recommend caution to observers who assume the loop plasma is isothermal, and hope that these results will influence the direction of coronal heating models and the efforts modelers spend on various heating scenarios.
\end{abstract}


\section{Introduction}

Recent results from solar X-ray and EUV imagers and spectrometers inspired a series of workshops which were designed to resolve the {\it coronal loop controversy}, where new observations were in conflict with the predictions of classical heating models. The model of Rosner et al. (1978), for example, assumed that a loop was single magnetic flux tube with no internal structure. The tube was filled with high-temperature, low-density plasma in hydrostatic equilibrium where the temperature could vary along but not across the field. These predictions were challenged by the results of Schmelz et al. (2001), who used spectral line data from the Coronal Diagnostics Spectrometer (CDS) on the Solar and Heliospheric Observatory and broadband data from the Soft X-ray Telescope (SXT) on Yohkoh to do Differential Emission Measure (DEM) analysis at several positions along their target loop. The resulting temperature distributions were clearly inconsistent with isothermal plasma in the cross-field direction. Subsequent analysis focusing on background subtraction (Schmelz et al. 2005; Schmelz \& Martens 2006), modeling (Martens et al. 2002; Weber et al. 2005), and similar results from other CDS loops (Cirtain et al. 2007; Schmelz et al 2007) supported the original claim that multithermal plasma was required to explain the CDS data. 

In parallel with the analysis described above, isothermal results continued to be published in the literature. Early papers used a simple ratio analysis, like the one employed by Kano \& Tsuneta (1996) for loops observed by SXT. This ratio analysis, however, $assumes$ an isothermal plasma, so obviously, the results presented in these papers could not be used as evidence of isothermal loop cross sections. Later analysis used DEM methods with strong high- and low-temperature constraints and data from the EUV Imaging Spectrometer (EIS) and the Atmospheric Imaging Assembly (AIA) to show that loops had both isothermal and multithermal cross sections (Warren et al. 2008; Schmelz et al. 2013a,b; Brooks et al. 2011, 2012). Some of these examples could be described as marginally multithermal, and others require a fairly broad DEM distribution (Schmelz et al. 2010b; 2011b). In fact, Schmelz et al. (2014a) found that the DEM width correlated with the DEM-weighted temperature where the cooler the loop, the narrower the DEM required to model the data.

Throughout this paper, we apply the definitions that were agreed upon by the participants of the first coronal loops workshop, which took place in Paris in 2002. A $loop$ is a distinct configuration in an observation and a $strand$ is an elementary flux tube. The observational results described above indicate that the loops analyzed in these various studies could not be a simple flux tube and that they must have some internal structure. These observations are consistent with the predictions of nanoflare coronal heating models (Cargill 1994; Cargill \& Klimchuk 1997), where unresolved strands reconnect and release small amounts of energy. Klimchuk (2009) describes the concept of a nanoflare storm. During the storm, bundles of unresolved strands are heated impulsively as reconnections release energy. {\bf But as the storm ends, reconnections cease, and since the cooling timescale tends to increase with decreasing temperature, hot strands cool relatively quickly and all strands spend more time cooling slowly through the lower coronal temperature range. Since there is now a preponderance of cool strands, the DEM is expected to narrow with time.} This effect was observed with AIA where the temperature evolution of a target loop was followed as it cooled. The DEM distribution required to model the data was broad early in the loop lifetime, but isothermal before the loop faded from view (Schmelz et al. 2014b).

{\bf Based on the cooling timescale argument outlined above, one might expect to find many more isothermal or marginally multithermal DEMs than broad DEMs. The ratio of these populations might reveal something about the heating timescale for nanoflare storms.} In an effort to determine which type of DEM distribution - isothermal or multithermal - dominates the loop populations, Schmelz et al. (2015) examined AR 11294, which was observed by AIA on 2011 September 15. In this original loop inventory paper, they examined the 171-\AA\ image of the target active region and selected all loop segments that were visible against a clean background. There were a variety of results: two segments were isothermal, six were effectively isothermal (a cooler AIA channel would be required for a definitive conclusion), one had both an isothermal transition region as well as a multithermal coronal solution, five required multithermal DEMs, and five could not be separated reliably from the background in other AIA channels. {\bf The rate of occurrence of a nanoflare on a strand must be faster than the cooling timescale to explain these results. To take an extreme example: if the timescale was very long then most of the plasma would be allowed to cool substantially and hence the DEMs would be more isothermal than multithermal.} In this paper, we continue the loop inventory analysis for six new regions. We hope that these results will affect the course of coronal heating models and the endeavors of modelers on diverse heating schemes.

\section{Observations}

The Solar Dynamics Observatory was launched on 2010 February 11 with the purpose of increasing the understanding of the solar magnetic field. It contains three instruments: AIA, the Extreme Ultraviolet Variability Explorer, and the Helioseismic and Magnetic Imager. The excellent resolution and cadence of the AIA allows for spectacular images of the Sun including a few wavelengths that were previously rarely observed. Specifically, the AIA has a field of view with 1.28 solar radii in the EW and NS directions and 0.6 arcsecond pixels, and the cadence is around 10 seconds. Of the ten filters in the AIA, observations were taken from the six filters centered on ionized iron including: Fe VIII (131 \AA), Fe IX (171 \AA), Fe XII (193 \AA), Fe XIV (211 \AA), Fe XVI (335 \AA), and Fe XVIII (94 \AA). {\bf Some filters contain flaring lines while others, e.g., the 94-\AA\ filter, contain lower-temperature lines. The former should not contribute to these observations, but the latter almost certainly do. Nevertheless, all of these lines are included in the appropriate response function (see below).} These six filters used in conjunction allow for both isothermal and multithermal analysis, which is a major boon to solar physics and, in particular, the study of the corona.  

Level 1 AIA data were obtained from the Virtual Solar Observatory\footnote{http://sdac.virtualsolar.org/cgi/search} website.  Using the AIA\_prep program included in SolarSoft\footnote{http://www.lmsal.com/solarsoft/} on the data then ensured proper alignment, rotation, and image scaling.  Occasionally when examining the images, it became apparent that a shift by a pixel in one filter would provide much better alignment with the other filters. 

In order to perform an effective temperature analysis, it is necessary to obtain a response function for each filter. {\bf To this end, the effective areas, the instrument platescale, and gain provided by the instrument teams were obtained from SolarSoft. A synthetic solar spectrum was constructed with atomic data and ionization equilibria from CHIANTI 7.1 (Dere et al. 1997; Landi et al. 2013) and the set of coronal element abundances from Schmelz et al. (2012).} The resulting response functions have units of DN s$^{-1}$ pixel$^{-1}$  per unit emission measure.

\section{Analysis}

Normally, when we select AIA loops for DEM analysis (e.g., Schmelz et al. 2011a,c), we would cycle through the images from all the coronal filters searching for targets. We would want these loops to be visible in at least three filters, the minimum required for DEM, and there would need to be a nearby area of reasonably clean background. The Loop Inventory process is a bit different, however. In this analysis, we are following the procedure outlined in the paper by Schmelz et al. (2015). Only the AIA 171-\AA\ images were used to select loop segments for detailed temperature analysis. There are several reasons for this. AIA results from the literature indicate that the loops seen in the 171-\AA\ images: (1) appear sharper and more readily discernable; (2) are likely to be cooling; and (3) are more prone to have isothermal cross-sections. This third point is especially important because, as described in the Introduction, our multithermal loop cross-sections led in part to the coronal loop controversy. These temperature results were not only new and unexpected, but also in conflict with the predictions of classical heating models. It was our responsibility to show that the data actually $required$ these more complex temperature solutions and to make every effort to understand why different types of observations were giving different results. In this work, we have done this due diligence by deliberately skewing our selection criteria toward the loop population that had traditionally shown isothermal cross-sections. {\bf Although we are not selecting out multithermal cases with this approach - our pilot study already shows that broad DEMs can have cool components - we are making every effort to incorporate isothermal cases in our sample. This criterion is consistent with the coronal cooling timescales, which as noted in the Introduction, increase with decreasing temperature, and with the results of Schmelz et al. (2014a,b), which show evidence for narrowing DEMs with decreasing temperature.}

AIA 171-\AA\ images of the six regions studied here are shown in Figure 1. Each small box marks a different loop segment, which was selected using the criteria described above. Table 1 lists the active regions, observation dates, solar coordinates, and the number of loop segments identified. Table 2 lists the individual loops with the same numbering scheme as the Figure. After these targets were selected, the other AIA filters were examined. We describe the appearance of each loop in each filter in Table 2.

Our data set reveals several examples of the simplest type of temperature analysis, where the loop is clearly visible in the AIA 171-\AA\ filter but not visible in the other filters. Figure 2 shows a $Rainbow\ Plot$ of Region A Loop 11 where the 131-\AA\ AIA image is shown in purple, 171 in blue, 193 in green, 211 in yellow, 225 in orange, and 94 in red. The position of the target, Loop 11, is designated with arrows. It is isothermal, within the temperature resolution of AIA, with Log T $\simeq$ 5.8, near the peak response of the 171-\AA\ AIA filter. These results are listed in the last two columns of Table 2.

With only a superficial examination of the Rainbow Plot in Figure 3, one might conclude that Region F Loop 7 is isothermal for reasons similar to those described in the last paragraph. There is an important difference, however. Although this loop is visible only in the 171-\AA\ image, it is not necessarily because the loop is not present in the other images. Rather, it sits in a crowded arcade and the background has come up in some of the other filters, possibly masking the appearance of the target with the increased emission measure of many unresolved structures. As a result, this loop and the others like it are not candidates for temperature analysis because they cannot be separated from the surrounding background. In the last two columns of Table 2, they are not assigned a temperature and are categorized as $Background$.

There are several examples of loops that we describe as $effectively$ isothermal where the plasma to which AIA is sensitive could not be distinguished from isothermal emission, within measurement uncertainties (e.g., Schmelz et al. 1996; 2014b). We illustrate this category with Region E Loop 9. Figure 4 shows that this loop is visible in the 131- and 171-\AA\ filters, but not visible in the hotter filters. This loop and other like it are categorized as effectively isothermal because although they could be multithermal at cool (T $<$ 1 MK) temperatures, we cannot know this without additional data. These temperatures can be found with ratio analysis, which we describe below.

The data in Table 2 indicate that there are many loops that are eligible for temperature analysis. With each candidate, 10 loop pixels and 10 background pixels are selected and averaged.  Once we calculated the standard deviations, we then subtracted the background and propagated the errors. These values were then normalized by the appropriate exposure time, resulting in units of $DN\ s^{-1}\ pixel^{-1}$ for each filter. If $I$ is the average intensity, then

\begin{equation}
I\ \propto\ \sum\ Resp(T)\ \times\ DEM(T)\ \Delta\ T , 
\end{equation}

\noindent
where Resp(T) is the instrument response function, DEM is the differential emission measure, and $T$ is the temperature. If the isothermal approximation applies, then we can pull the response out of the summation, and the equation simplifies to 

\begin{equation}
I\ \propto\ Resp(T)\ \times\ \sum\ DEM(T)\ \Delta\ T\ \propto\ Resp(T)\ \times\ EM .
\end{equation}

\noindent
$EM$ is the plasma emission measure, and we can use the ratio method to find the temperature:

\begin{equation}
{I_{131} \over I_{171}}  =  {{Resp_{131}(T)} \over {Resp_{171}(T)}}
\end{equation}

The ratio method is illustrated in Figure 5. The curve with the peak at Log T $=$ 5.55 is the 131-\AA\ response divided by the 171-\AA\ response. The flat lines show the observed intensity ratio (solid) and uncertainties (dashed). Each panel shows a different example from the data set. In general, the flat lines intersect the curve at two locations, both of which are candidates for the plasma temperature. These values and the associated uncertainties are listed in Table 2. In some cases, the solid (Region E Loop 07) and/or dashed (Region C Loop 13) line misses the curve, indicating that the data do not have a high enough signal-to-noise to pin down either the temperature value or the associated limit. {\bf These cases are indicated by a double dashed in Table 2.}

There are many examples of loops in our data sets that appear in multiple filters. The rainbow plot in Figure 6 is for Region B Loop 43, which is visible or barely visible in all the AIA coronal filters. For this and any loop segment that is visible in more than two filters, we cannot assume that the plasma is isothermal (although it may be {\bf if its temperature falls in a range where those filters have significant overlap}), and we cannot use the ratio method to find the temperature. For the analysis of these loops, we require DEM techniques. Schmelz et al. (2010a, 2011a,b,c) describe a method called DEM\_manual, which uses forward fitting. One option available in the DEM\_manual program is to find the best isothermal fit to the data. We used this option for all the loops that were visible in three or more filters. We first input a spike-shaped DEM at Log T $=$ 5.30, and the program determines the height of the spike that provides the best fit to the available data. We then move to the next temperature bin and repeat the process. In general, for quiescent loops, we continue to Log T $=$ 6.60, but can go higher for flares. 

The best isothermal fit is the spike DEM with the lowest reduced ${\chi^2}$. The results for several different loops are shown in Figure 7, where the blue spike is the best isothermal fit and the reduced ${\chi^2}$, also in blue, is listed in the upper right corner. The predicted-to-observed intensity ratios for these DEM\_manual results are plotted as open diamonds in Figure 8. We use the Rainbow Plot color coding which corresponds to the temperature sequence for these loops: 131 (purple), 171 (blue), 193 (green), 211 (yellow), 225 (orange), and 94 (red). In some cases, this method finds a good fit to the data with reduced ${\chi^2}\ <\ 1$. One example is for Region E Loop 25, which is plotted in the first panel of Figures 7 and 8. The loop is visible in the three coolest coronal filters, and the analysis done with DEM\_manual shows that isothermal plasma can reproduce the observations, within the uncertainties. This loop is effectively isothermal (since it is visible in the 131-\AA\ filter), and the resulting temperature is recorded in Table 2.

Many cases, however, show that even the best isothermal result is not a good fit to the data. The reduced ${\chi^2}$ values in Figure 7 are too high and the predicted-to-observed intensity ratios for some filters in Figure 8 are significantly different from one. For these cases, we use XRT\_dem\_iterative2 (Weber et al. 2004; Schmelz et al. 2009) to generate 100 multithermal Monte-Carlo realizations. These are shown in red in the panels of Figure 7. The black curve is the best fit and the corresponding reduced ${\chi^2}$ is listed in the upper right corner, also in black. In these cases, the DEM-weighted temperature is listed in Table 2. The predicted-to-observed intensity ratios for the best-fit results from XRT\_dem\_iterative2 are shown as filled squares in Figure 8.

\newcommand{\ltapprox}{\stackrel{<}{\sim}}

\section{Discussion}

The results of our analysis discussed in the previous section and summarized in Table 2 support and expand the findings of the loop inventory pilot study of Schmelz et al. (2015). We have added 188 targets to the original 19 from the pilot. The most common entry in the comments column of Table 2 is $background$, which indicates that the target loop cannot be reliably separated from the background in one or more of the AIA coronal channels required for temperature analysis. This occurred in 86 of the loops in Table 2 and five from the pilot study. We can therefore confirm earlier results that separating the loop from the coronal background represents a serious and, in many cases, insurmountable challenge, {\bf although some improvement might be expected using observations made with a spectrometer.} This finding comes with an associated caution to other researchers. Although there may be a distinct AIA 171-\AA\ loop segment, it could be part of an arcade that gets more densely populated with increasing temperature. The target may indeed be present in the 193- and 211-\AA\ images, but the background may be too dense to see it (see, e.g., Brickhouse \& Schmelz 2006). An unintended result of an automatic background subtraction algorithm might be that a high background might be subtracted from the embedded loop leaving essentially zero flux. This is one way to misidentify loops as isothermal rather than multithermal.

The next most common entry in Table 2 is $multithermal$. This is despite trying to bias our selection criteria toward loops with isothermal cross sections. Our 61 examples here can be added to the original five from the pilot study. These are the loop segments that appeared in at least three of the AIA coronal filters and where the background subtracted intensities could not be reproduced with an isothermal model, within uncertainties. We ran both DEM\_manual and XRT\_dem\_iterative2 on these data and got good results, with small reduced ${\chi^2}$ values. The DEM-weighted temperature for these examples is listed in the last column of Table 2.

The third most common entry in Table 2 is $effectively\ isothermal$. This tally includes 28 loops from this study plus six from the pilot. This descriptive indicates that the emission to which AIA is sensitive could not be distinguished from isothermal emission, within measurement uncertainties. These loops may, however, be multithermal at transition region temperatures, but we would require data from a different instrument, perhaps the EUV Imaging Spectrometer or the Interface Region Imaging Spectrograph, to determine the true cross-field temperature distribution of these examples. We encourage observers with expertise in the analysis of data from these instruments to look into this question. 

Ten entries from this study join two from the pilot in the $isothermal$ category from Table 2. These are cases where the loop is visible only at 171-\AA\ and there is clean background in 131- and 193-\AA\ images. Our survey also appears to have caught one rather small flare, Region C Loop 6, which was not visible in GOES data. DEM analysis was done on these data, and a good result was obtained. We also identified one loop with so much transition-region temperature plasma that the AIA data alone were not able to constrain the DEM model, Region B Loop 35. This loop is labeled $unconstrained$ in the comments column of Table 2. We complete our survey with a single $inconclusive$ entry, Region D Loop 12. Despite multiple attempts at pixel selection for both the loop and the background, we were not able to find a DEM model that successfully reproduced the fluxes. One possible explanation for this might be that there was something behind our loop that we were not resolving properly.

Our survey can confirm an unexpected finding from the pilot study. We also found no loop segments where we could assume the plasma was isothermal and properly use the 171-to-193 ratio method, which would be similar to the analysis done for many loops observed with TRACE and EIT. This would require loops to be visible in the 171- and 193-\AA\ images only with clean background in the 131- and 211-\AA\ images. In our examples and in those of the pilot study, the 171-\AA\ segment is either not visible in the 193-\AA\ filter, masked by complex background in the 193-\AA\ filter, or visible in the 193-\AA\ filter but also visible in other filters. The size of the present study indicate that this is probably not just a coincidence, and is more likely to reflect how the loops are cooling. 

{\bf All of the isothermal DEMs identified here are below 1 MK. The plasma starts to cool more quickly below this temperature as the peak of the radiative loss curve is reached. Consequently, one may expect to see fewer purely isothermal DEMs/loop segments simply because the plasma cools relatively quickly below 1 MK and so there are fewer examples to be observed. It could also be that plasma tends to be re-heated before the DEMs approach isothermality.} Evidence of a correlation between the DEM-weighted temperature and the cross-field DEM width for coronal loops was found by Schmelz et al. (2014a). Their AIA, XRT, and EIS data showed that warmer loops require broader DEMs. A target loop analyzed by Schmelz et al. (2014b), which was cooling through the AIA passbands, was also evolving from a broad DEM to a narrow DEM. 
{\bf These results are consistent with the radiative loss function in the coronal temperature range.} If loops are indeed composed of bundles of unresolved magnetic strands, then our results could indicate that fewer strands are emitting in the later cooling phase, consistent with the nanoflare storm model (Klimchuk 2009), and potentially giving us more insight into the long standing isothermal $versus$ multithermal component of the coronal loop controversy.

\section{Conclusions}

We have expanded the results of the pilot study done by Schmelz et al. (2015) to include multiple regions. The original Loop Inventory included only one active region, AR 11294, and 19 loops. We have added 188 loops to the analysis. In both the pilot as well as the current study, we examined the AIA 171-\AA\ images and selected all loop segments with a clean background. Many of these targets, five in the pilot and 86 here, could not be separated from the coronal background in higher temperature images, and were therefore, not candidates for temperature analysis. Five loops from the pilot and 61 here were visible in three or more AIA filters, and DEM analysis revealed that they had multithermal cross sections. Six loops from the pilot and 28 here were effectively isothermal, and two plus 10 were isothermal. The pilot had one loop that had both isothermal and multithermal solutions. The current work included one small flare, one example where the data available could not constrain the DEM, and one example where the results were inconclusive. 

The results of our inventory indicate that even a loop seen in a 171-\AA\ image is significantly more likely than not to have a multithermal cross section. This implies that both observers and modelers should exercise caution when making simplistic assumptions about the temperature structure of coronal loops. The data may require DEM methods, which are much more difficult to use than simplistic ratio techniques. These results should also influence the direction of coronal heating models and the efforts that models spend investigating the fundamental properties of loops. 

\acknowledgements

This work was inspired by discussions at the Coronal Loops Workshops. The Atmospheric Imaging Assembly on the Solar Dynamics Observatory is part of NASA's Living with a Star program. CHIANTI is a collaborative project involving the NRL (USA), the Universities of Florence (Italy) and Cambridge (UK), and George Mason University (USA). Solar physics research at the University of Memphis is supported by a $Hinode$ subcontract from NASA/SAO. 

{}

\clearpage

\begin{deluxetable}{lllll}
\tabletypesize{\scriptsize}
\tablewidth{0pt}
\renewcommand{\tabcolsep}{5.0pt}
\tablecaption{AIA Active Regions}
\tablehead{
\colhead{ } & \colhead{Region} & \colhead{Date} & \colhead{Coordinates} & \colhead{\# Loops} 
}
\startdata
A	&	11158	&	2011 February 15	&	S21 W28	&	37	\\
B	&	11166	&	2011 March 06	&	N10 E27	&	51	\\
C	&	11330 (east)	&2011 October 27	&	N13 E13	&	13	\\
D	&	11330 (west)	&2011 October 27	&	N06 E01 	&	30	\\
E	&	11944 (east)	&2014 January 06	&	S11 E22 	&	26	\\
F	&	11944 (west)	&2014 January 06	&	S08 E08	&	31
\enddata
\end{deluxetable} 

\clearpage

\begin{deluxetable}{lllllllllllll}
\tabletypesize{\scriptsize}
\rotate
\tablewidth{0pt}
\renewcommand{\tabcolsep}{5.0pt}
\tablecaption{AIA Loop Segments and Properties}
\tablehead{
\colhead{\# } & \colhead{131} & \colhead{171} & \colhead{193} & \colhead{211} & \colhead{335} & \colhead{94} & \colhead{Comments} & \colhead{Temp}
}
\startdata
\hline																	\\
Region A																	\\
\hline																	\\
1	&	visible	&	visible	&	visible	&	visible	&	visible	&	background	&	multithermal	&	6.06	\\
2	&	visible	&	visible	&	visible	&	visible	&	background	&	background	&	multithermal	&	5.96	\\
3	&	background	&	visible	&	background	&	background	&	background	&	background	&	background	&		\\
4	&	barely visible	&	visible	&	visible	&	visible	&	background	&	background	&	multithermal	&	6.25	\\
5	&	barely visible	&	visible	&	visible	&	visible	&	background	&	background	&	multithermal	&	6.38	\\
6	&	visible	&	visible	&	background	&	background	&	background	&	background	&	background	&		\\
7	&	visible	&	visible	&	background	&	background	&	background	&	background	&	background	&		\\
8	&	background	&	visible	&	background	&	background	&	background	&	background	&	background	&		\\
9	&	not visible	&	visible	&	not visible	&	not visible	&	not visible	&	not visible	&	isothermal	&	5.80	\\
10	&	visible	&	visible	&	visible	&	visible	&	not visible	&	not visible	&	multithermal	&	6.20	\\
11	&	not visible	&	visible	&	not visible	&	not visible	&	not visible	&	not visible	&	isothermal	&	5.80	\\
12	&	visible	&	visible	&	visible	&	visible	&	barely visible	&	not visible	&	multithermal	&	6.12	\\
13	&	not visible	&	visible	&	visible	&	background	&	not visible	&	not visible	&	background	&		\\
14	&	barely visible	&	visible	&	visible	&	visible	&	background	&	not visible	&	multithermal	&	6.30	\\
15	&	not visible	&	visible	&	visible	&	visible	&	background	&	not visible	&	background	&		\\
16	&	not visible	&	visible	&	background	&	background	&	background	&	not visible	&	background	&		\\
17	&	not visible	&	visible	&	background	&	background	&	not visible	&	not visible	&	background	&		\\
18	&	not visible	&	visible	&	not visible	&	not visible	&	background	&	background	&	isothermal	&	5.80	\\
19	&	not visible	&	visible	&	not visible	&	not visible	&	not visible	&	not visible	&	isothermal	&	5.80	\\
20	&	visible	&	visible	&	visible	&	barely visible	&	background	&	background	&	multithermal	&	6.31	\\
21	&	visible	&	visible	&	not visible	&	not visible	&	not visible	&	not visible	&	eff isothermal	&	5.43$^{+.01}_{-.02}$, 5.78$^{+.04}_{-.03}$	\\
22	&	background	&	visible	&	visible	&	visible	&	background	&	background	&	multithermal	&	6.98	\\
23	&	background	&	visible	&	visible	&	visible	&	visible	&	background	&	multithermal	&	6.64	\\
24	&	not visible	&	visible	&	not visible	&	not visible	&	not visible	&	not visible	&	isothermal	&	5.80	\\
25	&	not visible	&	visible	&	background	&	not visible	&	not visible	&	not visible	&	background	&		\\
26	&	barely visible	&	visible	&	background	&	not visible	&	not visible	&	not visible	&	background	&		\\
27	&	barely visible	&	visible	&	background	&	background	&	not visible	&	not visible	&	background	&		\\
28	&	visible	&	visible	&	background	&	background	&	not visible	&	not visible	&	background	&		\\
29	&	background	&	visible	&	background	&	background	&	not visible	&	not visible	&	background	&		\\
30	&	background	&	visible	&	background	&	background	&	not visible	&	not visible	&	background	&		\\
31	&	visible	&	visible	&	background	&	barely visible	&	not visible	&	not visible	&	background	&		\\
32	&	not visible	&	visible	&	background	&	background	&	not visible	&	not visible	&	background	&		\\
33	&	not visible	&	visible	&	background	&	not visible	&	not visible	&	not visible	&	background	&		\\
34	&	not visible	&	visible	&	background	&	not visible	&	not visible	&	not visible	&	background	&		\\
35	&	barely visible	&	visible	&	barely visible	&	background	&	not visible	&	not visible	&	multithermal	&	6.32	\\
36	&	not visible	&	visible	&	visible	&	background	&	not visible	&	not visible	&	background	&		\\
37	&	background	&	visible	&	not visible	&	not visible	&	not visible	&	not visible	&	background	&		\\
\hline																	\\
Region B																	\\
\hline																	\\
1	&	visible	&	visible	&	background	&	not visible	&	not visible	&	not visible	&	background	&		\\
2	&	visible	&	visible	&	not visible	&	not visible	&	not visible	&	not visible	&	eff isothermal	&	5.38$^{+.02}_{-.04}$, 5.88$^{+.05}_{-.04}$	\\
3	&	not visible	&	visible	&	not visible	&	not visible	&	not visible	&	not visible	&	isothermal	&	5.80	\\
4	&	not visible	&	visible	&	not visible	&	not visible	&	not visible	&	not visible	&	isothermal	&	5.80	\\
5	&	visible	&	visible	&	visible	&	visible	&	not visible	&	barely visible	&	multithermal	&	6.05	\\
6	&	visible	&	visible	&	visible	&	background	&	not visible	&	background	&	background	&		\\
7	&	barely visible	&	visible	&	background	&	not visible	&	not visible	&	not visible	&	background	&		\\
8	&	visible	&	visible	&	visible	&	visible	&	background	&	background	&	multithermal	&	6.01	\\
9	&	barely visible	&	visible	&	visible	&	visible	&	not visible	&	not visible	&	multithermal	&	6.16	\\
10	&	not visible	&	visible	&	not visible	&	not visible	&	not visible	&	not visible	&	isothermal	&	5.80	\\
11	&	visible	&	visible	&	not visible	&	not visible	&	not visible	&	not visible	&	eff isothermal	&	5.37$^{+.01}_{-.02}$, 5.91$^{+.05}_{-.03}$	\\
12	&	visible	&	visible	&	visible	&	visible	&	visible	&	barely visible	&	multithermal	&	5.96	\\
13	&	visible	&	visible	&	barely visible	&	not visible	&	not visible	&	background	&	eff isothermal	&	6.40	\\
14	&	barely visible	&	visible	&	not visible	&	not visible	&	not visible	&	not visible	&	eff isothermal	&	5.41$^{+.03}_{-.07}$, 5.81$^{+.16}_{-.06}$	\\
15	&	barely visible	&	visible	&	not visible	&	not visible	&	not visible	&	not visible	&	eff isothermal	&	5.41$^{+.03}_{-.07}$, 5.81$^{+.16}_{-.06}$	\\
16	&	background	&	visible	&	visible	&	background	&	background	&	background	&	background	&		\\
17	&	background	&	visible	&	background	&	not visible	&	not visible	&	not visible	&	background	&		\\
18	&	background	&	visible	&	background	&	background	&	background	&	background	&	background	&		\\
19	&	visible	&	visible	&	visible	&	visible	&	visible	&	visible	&	multithermal	&	6.33	\\
20	&	not visible	&	visible	&	background	&	background	&	background	&	background	&	background	&		\\
21	&	visible	&	visible	&	not visible	&	not visible	&	not visible	&	not visible	&	background	&		\\
22	&	background	&	visible	&	background	&	background	&	not visible	&	background	&	background	&		\\
23	&	barely visible	&	visible	&	not visible	&	not visible	&	not visible	&	background	&	eff isothermal	&	5.35$^{+.01}_{-.04}$, 5.96$^{+.05}_{-.06}$	\\
24	&	not visible	&	visible	&	background	&	background	&	not visible	&	not visible	&	background	&		\\
25	&	barely visible	&	visible	&	not visible	&	not visible	&	background	&	background	&	eff isothermal	&	5.35$^{+.01}_{-.04}$, 5.96$^{+.05}_{-.06}$	\\
26	&	barely visible	&	visible	&	background	&	background	&	background	&	background	&	background	&		\\
27	&	background	&	visible	&	not visible	&	not visible	&	not visible	&	not visible	&	background	&		\\
28	&	barely visible	&	visible	&	background	&	background	&	not visible	&	background	&	background	&		\\
29	&	barely visible	&	visible	&	barely visible	&	background	&	not visible	&	not visible	&	multithermal	&	6.09	\\
30	&	background	&	visible	&	not visible	&	not visible	&	not visible	&	not visible	&	background	&		\\
31	&	visible	&	visible	&	not visible	&	not visible	&	not visible	&	not visible	&	eff isothermal	&	5.36$^{+.02}_{-.06}$, 5.93$^{+.17}_{-.06}$	\\
32	&	visible	&	visible	&	visible	&	visible	&	background	&	not visible	&	multithermal	&	6.45	\\
33	&	not visible	&	visible	&	not visible	&	not visible	&	not visible	&	not visible	&	isothermal	&	5.80	\\
34	&	not visible	&	visible	&	visible	&	background	&	not visible	&	not visible	&	background	&		\\
35	&	barely visible	&	visible	&	visible	&	barely visible	&	background	&	background	&	unconstrained	&		\\
36	&	not visible	&	visible	&	visible	&	barely visible	&	not visible	&	not visible	&	multithermal	&	6.14	\\
37	&	visible	&	visible	&	visible	&	visible	&	barely visible	&	background	&	multithermal	&	6.70	\\
38	&	background	&	visible	&	not visible	&	not visible	&	not visible	&	not visible	&	background	&		\\
39	&	barely visible	&	visible	&	visible	&	visible	&	background	&	not visible	&	multithermal	&	6.33	\\
40	&	barely visible	&	visible	&	background	&	background	&	background	&	not visible	&	background	&		\\
41	&	barely visible	&	visible	&	not visible	&	not visible	&	not visible	&	not visible	&	eff isothermal	&	5.37$^{+.03}_{-.05}$, 5.90$^{+.11}_{-.06}$	\\
42	&	not visible	&	visible	&	visible	&	barely visible	&	not visible	&	not visible	&	multithermal	&	6.00	\\
43	&	visible	&	visible	&	visible	&	visible	&	visible	&	barely visible	&	multithermal	&	5.91	\\
44	&	visible	&	visible	&	barely visible	&	barely visible	&	background	&	not visible	&	eff isothermal 	&	5.70	\\
45	&	background	&	visible	&	visible	&	barely visible	&	barely visible	&	not visible	&	background	&		\\
46	&	visible	&	visible	&	not visible	&	not visible	&	not visible	&	not visible	&	eff isothermal	&	5.38$^{+.03}_{-.11}$, 5.88$^{+--}_{-.06}$	\\
47	&	barely visible	&	visible	&	visible	&	barely visible	&	background	&	background	&	multithermal	&	6.60	\\
48	&	visible	&	visible	&	background	&	not visible	&	not visible	&	not visible	&	background	&		\\
49	&	visible	&	visible	&	visible	&	visible	&	barely visible	&	not visible	&	multithermal	&	5.98	\\
50	&	visible	&	visible	&	visible	&	barely visible	&	barely visible	&	background	&	multithermal	&	6.49	\\
51	&	visible	&	visible	&	visible	&	background	&	not visible	&	not visible	&	background	&		\\
\hline																	\\
Region C																	\\
\hline																	\\
1	&	visible	&	visible	&	visible	&	barely visible	&	background	&	not visible	&	background	&		\\
2	&	visible	&	visible	&	visible	&	background	&	background	&	not visible	&	multithermal	&	6.14	\\
3	&	visible	&	visible	&	not visible	&	not visible	&	not visible	&	not visible	&	eff isothermal	&	5.35$^{+.01}_{-.04}$, 5.96$^{+.05}_{-.06}$	\\
4	&	barely visible	&	visible	&	background	&	background	&	background	&	background	&	background	&		\\
5	&	visible	&	visible	&	visible	&	visible	&	background	&	not visible	&	multithermal	&	6.28	\\
6	&	visible	&	visible	&	visible	&	visible	&	visible	&	background	&	small flare	&	6.74	\\
7	&	visible	&	visible	&	visible	&	visible	&	background	&	background	&	multithermal	&	6.60	\\
8	&	visible	&	visible	&	visible	&	barely visible	&	not visible	&	not visible	&	multithermal	&	6.01	\\
9	&	not visible	&	visible	&	background	&	not visible	&	not visible	&	not visible	&	background	&		\\
10	&	background	&	visible	&	barely visible	&	not visible	&	not visible	&	not visible	&	background	&		\\
11	&	background	&	visible	&	visible	&	barely visible	&	not visible	&	not visible	&	background	&		\\
12	&	not visible	&	visible	&	background	&	background	&	not visible	&	not visible	&	background	&		\\
13	&	barely visible	&	visible	&	not visible	&	not visible	&	not visible	&	not visible	&	eff isothermal	&	5.32$^{+.05}_{---}$, 6.00$^{+--}_{-.10}$	\\
\hline																	\\
Region D																	\\
\hline																	\\
1	&	visible	&	visible	&	visible	&	visible	&	background	&	background	&	multithermal	&	6.22	\\
2	&	visible	&	visible	&	visible	&	background	&	background	&	barely visible	&	multithermal	&	6.13	\\
3	&	visible	&	visible	&	visible	&	background	&	not visible	&	background	&	background	&		\\
4	&	barely visible	&	visible	&	background	&	background	&	background	&	background	&	background	&		\\
5	&	visible	&	visible	&	visible	&	background	&	background	&	barely visible	&	multithermal	&	6.30	\\
6	&	visible	&	visible	&	visible	&	background	&	background	&	not visible	&	multithermal	&	6.11	\\
7	&	visible	&	visible	&	visible	&	visible	&	barely visible	&	visible	&	multithermal	&	6.20	\\
8	&	visible	&	visible	&	visible	&	visible	&	barely visible	&	visible	&	multithermal	&	6.50	\\
9	&	visible	&	visible	&	visible	&	background	&	not visible	&	not visible	&	multithermal	&	6.22	\\
10	&	background	&	visible	&	visible	&	background	&	background	&	background	&	background	&		\\
11	&	visible	&	visible	&	visible	&	visible	&	visible	&	visible	&	multithermal	&	6.30	\\
12	&	barely visible	&	visible	&	visible	&	visible	&	barely visible	&	barely visible	&	inconclusive	&		\\
13	&	barely visible	&	visible	&	visible	&	background	&	background	&	background	&	background	&		\\
14	&	visible	&	visible	&	visible	&	visible	&	visible	&	visible	&	multithermal	&	6.05	\\
15	&	background	&	visible	&	background	&	background	&	background	&	background	&	background	&		\\
16	&	visible	&	visible	&	visible	&	visible	&	background	&	visible	&	multithermal	&	6.21	\\
17	&	barely visible	&	visible	&	visible	&	visible	&	not visible	&	background	&	multithermal	&	6.26	\\
18	&	barely visible	&	visible	&	visible	&	visible	&	background	&	barely visible	&	multithermal	&	6.18	\\
19	&	visible	&	visible	&	visible	&	visible	&	background	&	not visible	&	multithermal	&	6.18	\\
20	&	visible	&	visible	&	visible	&	barely visible	&	background	&	barely visible	&	background	&		\\
21	&	visible	&	visible	&	visible	&	visible	&	background	&	background	&	multithermal	&	6.06	\\
22	&	background	&	visible	&	background	&	background	&	background	&	background	&	background	&		\\
23	&	visible	&	visible	&	visible	&	background	&	background	&	background	&	background	&		\\
24	&	background	&	visible	&	not visible	&	not visible	&	not visible	&	background	&	background	&		\\
25	&	visible	&	visible	&	background	&	not visible	&	not visible	&	not visible	&	background	&		\\
26	&	visible	&	visible	&	visible	&	visible	&	visible	&	barely visible	&	multithermal	&	5.96	\\
27	&	barely visible	&	visible	&	barely visible	&	background	&	not visible	&	not visible	&	background	&		\\
28	&	background	&	visible	&	barely visible	&	barely visible	&	background	&	not visible	&	background	&		\\
29	&	background	&	visible	&	background	&	background	&	background	&	background	&	background	&		\\
30	&	background	&	visible	&	background	&	not visible	&	not visible	&	background	&	background	&		\\
\hline																	\\
Region E																	\\
\hline																	\\
1	&	background	&	visible	&	background	&	background	&	background	&	not visible	&	background	&		\\
2	&	visible	&	visible	&	not visible	&	not visible	&	not visible	&	barely visible	&	eff isothermal	&	6.10	\\
3	&	barely visible	&	visible	&	background	&	not visible	&	not visible	&	not visible	&	background	&		\\
4	&	background	&	visible	&	barely visible	&	not visible	&	not visible	&	not visible	&	background	&		\\
5	&	not visible	&	visible	&	not visible	&	not visible	&	not visible	&	not visible	&	isothermal	&	5.80	\\
6	&	not visible	&	visible	&	background	&	not visible	&	not visible	&	not visible	&	background	&		\\
7	&	barely visible	&	visible	&	not visible	&	not visible	&	not visible	&	background	&	eff isothermal	&	5.28$^{+.09}_{---}$, -.--$^{+--}_{---}$	\\
8	&	visible	&	visible	&	visible	&	background	&	not visible	&	background	&	multithermal	&	6.11	\\
9	&	visible	&	visible	&	not visible	&	not visible	&	not visible	&	background	&	eff isothermal	&	5.35$^{+.02}_{-.07}$, 5.94$^{+--}_{-.07}$	\\
10	&	visible	&	visible	&	visible	&	barely visible	&	background	&	background	&	background	&	6.05	\\
11	&	visible	&	visible	&	visible	&	barely visible	&	background	&	background	&	multithermal	&	6.06	\\
12	&	barely visible	&	visible	&	not visible	&	not visible	&	not visible	&	background	&	eff isothermal	&	5.36$^{+.01}_{-.02}$, 5.93$^{+.05}_{-.05}$	\\
13	&	barely visible	&	visible	&	visible	&	barely visible	&	not visible	&	background	&	multithermal	&	6.07	\\
14	&	visible	&	visible	&	not visible	&	not visible	&	not visible	&	background	&	eff isothermal	&	5.35$^{+.03}_{-.03}$, 5.95$^{+.10}_{-.04}$	\\
15	&	visible	&	visible	&	visible	&	visible	&	background	&	background	&	multithermal	&	6.06	\\
16	&	background	&	visible	&	background	&	background	&	not visible	&	background	&	background	&		\\
17	&	visible	&	visible	&	visible	&	visible	&	visible	&	visible	&	multithermal	&	6.05	\\
18	&	barely visible	&	visible	&	visible	&	background	&	background	&	barely visible	&	multithermal	&	6.29	\\
19	&	background	&	visible	&	background	&	background	&	background	&	background	&	background	&		\\
20	&	barely visible	&	visible	&	not visible	&	not visible	&	not visible	&	not visible	&	eff isothermal	&	5.36$^{+.02}_{-.21}$, 5.93$^{+--}_{-.06}$	\\
21	&	barely visible	&	visible	&	not visible	&	not visible	&	not visible	&	not visible	&	eff isothermal	&	5.37$^{+.03}_{-.02}$, 5.88$^{+.08}_{-.04}$	\\
22	&	barely visible	&	visible	&	background	&	not visible	&	not visible	&	not visible	&	background	&		\\
23	&	visible	&	visible	&	visible	&	visible	&	visible	&	visible	&	multithermal	&	6.15	\\
24	&	barely visible	&	visible	&	not visible	&	not visible	&	not visible	&	background	&	eff isothermal	&	5.36$^{+.02}_{-.34}$, 5.94$^{+--}_{-.07}$	\\
25	&	visible	&	visible	&	barely visible	&	not visible	&	not visible	&	not visible	&	eff isothermal	&	6.05	\\
26	&	background	&	visible	&	background	&	background	&	background	&	background	&	background	&		\\
\hline																	\\
Region F																	\\
\hline																	\\
1	&	visible	&	visible	&	visible	&	barely visible	&	barely visible	&	barely visible	&	multithermal	&	5.93	\\
2	&	barely visible	&	visible	&	background	&	not visible	&	not visible	&	background	&	background	&		\\
3	&	visible	&	visible	&	visible	&	visible	&	background	&	background	&	multithermal	&	6.04	\\
4	&	visible	&	visible	&	background	&	background	&	not visible	&	not visible	&	background	&		\\
5	&	barely visible	&	visible	&	not visible	&	not visible	&	not visible	&	not visible	&	eff isothermal	&	5.35$^{+.03}_{-.30}$, 5.94$^{+--}_{-.07}$	\\
6	&	background	&	visible	&	visible	&	visible	&	background	&	background	&	multithermal	&	6.14	\\
7	&	background	&	visible	&	not visible	&	not visible	&	not visible	&	not visible	&	background	&		\\
8	&	barely visible	&	visible	&	not visible	&	not visible	&	not visible	&	not visible	&	eff isothermal	&	5.37$^{+.03}_{-.04}$, 5.90$^{+.10}_{-.05}$	\\
9	&	barely visible	&	visible	&	background	&	background	&	not visible	&	not visible	&	background	&		\\
10	&	barely visible	&	visible	&	barely visible	&	barely visible	&	barely visible	&	not visible	&	multithermal	&	6.31	\\
11	&	visible	&	visible	&	visible	&	barely visible	&	background	&	background	&	background	&		\\
12	&	background	&	visible	&	background	&	background	&	not visible	&	not visible	&	background	&		\\
13	&	visible	&	visible	&	visible	&	barely visible	&	background	&	not visible	&	background	&		\\
14	&	barely visible	&	visible	&	background	&	not visible	&	not visible	&	background	&	background	&		\\
15	&	visible	&	visible	&	 visible	&	 visible	&	background	&	not visible	&	multithermal	&	6.01	\\
16	&	visible	&	visible	&	visible	&	visible	&	visible	&	 visible	&	multithermal	&	6.50	\\
17	&	barely visible	&	visible	&	background	&	background	&	not visible	&	not visible	&	background	&		\\
18	&	background	&	visible	&	background	&	background	&	background	&	background	&	background	&		\\
19	&	visible	&	visible	&	not visible	&	not visible	&	not visible	&	not visible	&	eff isothermal	&	5.35$^{+.02}_{-.15}$, 5.96$^{+--}_{-.13}$	\\
20	&	barely visible	&	visible	&	visible	&	visible	&	barely visible	&	visible	&	multithermal	&	6.07	\\
21	&	visible	&	visible	&	visible	&	visible	&	background	&	background	&	multithermal	&	6.38	\\
22	&	barely visible	&	visible	&	background	&	background	&	background	&	background	&	background	&		\\
23	&	background	&	visible	&	visible	&	barely visible	&	background	&	background	&	background	&		\\
24	&	barely visible	&	visible	&	visible	&	barely visible	&	background	&	background	&	background	&	6.05	\\
25	&	background	&	visible	&	not visible	&	not visible	&	not visible	&	not visible	&	background	&		\\
26	&	barely visible	&	visible	&	background	&	background	&	not visible	&	not visible	&	background	&		\\
27	&	visible	&	visible	&	background	&	not visible	&	not visible	&	not visible	&	background	&		\\
28	&	visible	&	visible	&	not visible	&	not visible	&	not visible	&	not visible	&	eff isothermal	&	5.36$^{+.02}_{-.05}$, 5.92$^{+.15}_{-.05}$	\\
29	&	background	&	visible	&	background	&	background	&	background	&	background	&	background	&		\\
30	&	visible	&	visible	&	not visible	&	not visible	&	not visible	&	not visible	&	eff isothermal	&	5.37$^{+.02}_{-.02}$, 5.89$^{+.07}_{-.03}$	\\
31	&	background	&	visible	&	visible	&	visible	&	not visible	&	not visible	&	multithermal	&	6.20	
\enddata
\end{deluxetable} 

\clearpage

\begin{figure}
\epsscale{.7}
\plotone{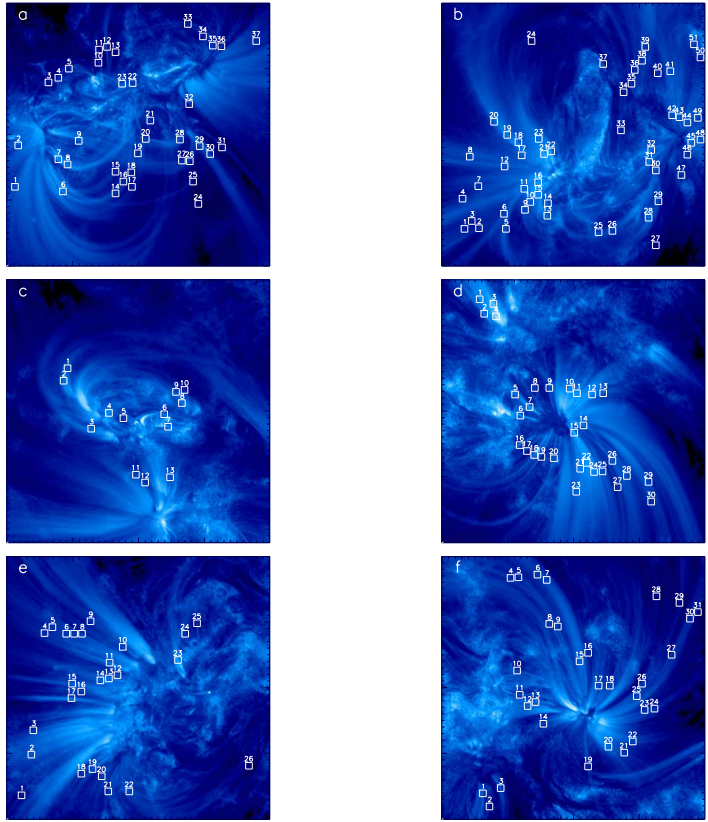}
\caption{AIA 171-\AA\ filter images of the active regions described in Table 1. Each loop segment is selected based on criteria described in the text and designated with a small white box and number. The main ion contributing to this filter is Fe IX with a peak formation temperature of Log T = 5.8
}
\end{figure}

\clearpage

\begin{figure}
\epsscale{.9}
\plotone{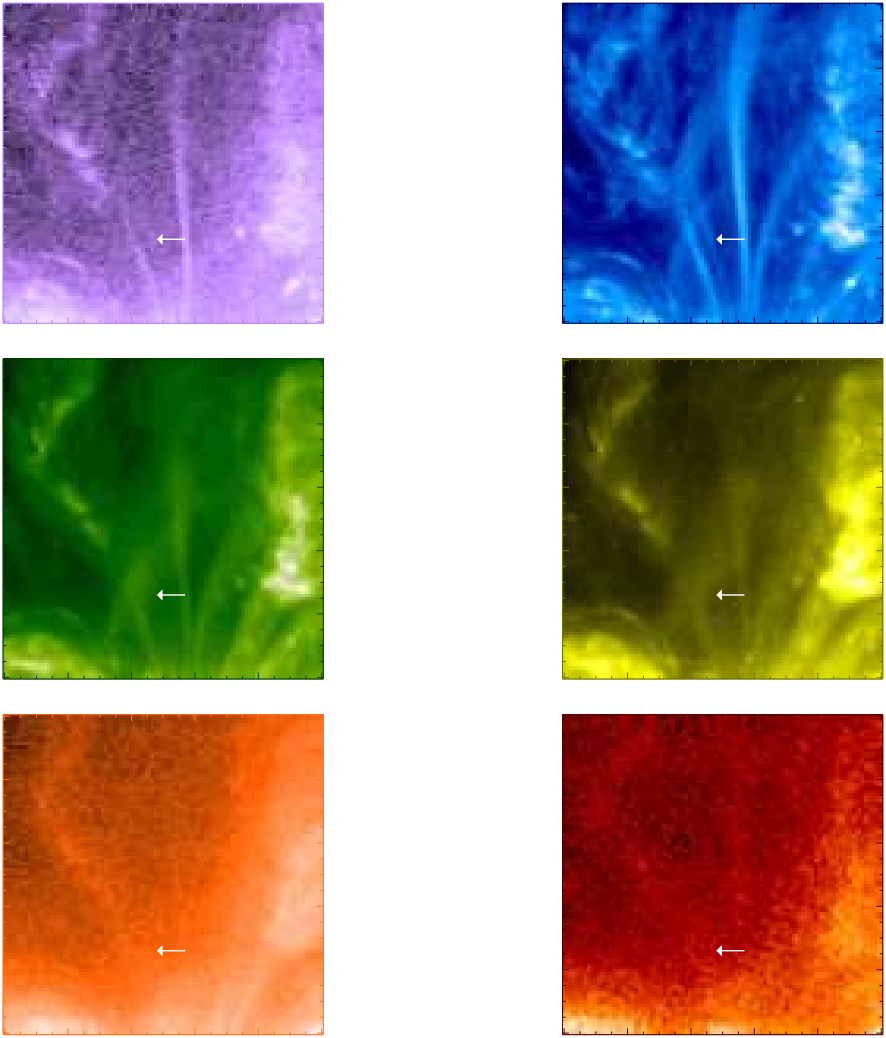}
\caption{Rainbow Plot with close-up field-of-view showing the subsection of AR 11158 containing Region A Loop 11. The image from the 131-\AA\ filter is shown in purple, 171 in blue, 193 in green, 211 in yellow, 335 in orange, and 94 in red. The position of the target, Loop 11, is designated with arrows. Loop 11 is an example of an isothermal loop.
}
\end{figure}

\clearpage

\begin{figure}
\epsscale{.9}
\plotone{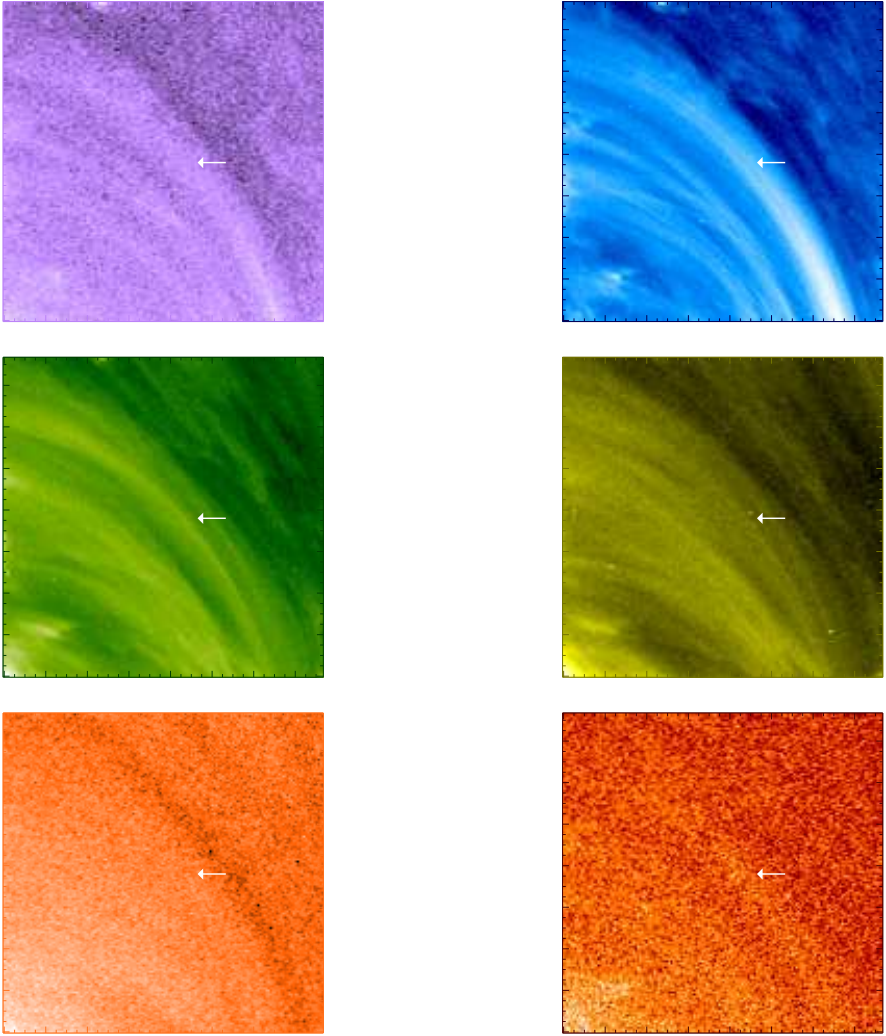}
\caption{Rainbow Plot with a format similar to Figure 2 showing the subsection of AR 11944 containing Region F Loop 7. Loop 7 is categorized as Background.
}
\end{figure}

\clearpage

\begin{figure}
\epsscale{.9}
\plotone{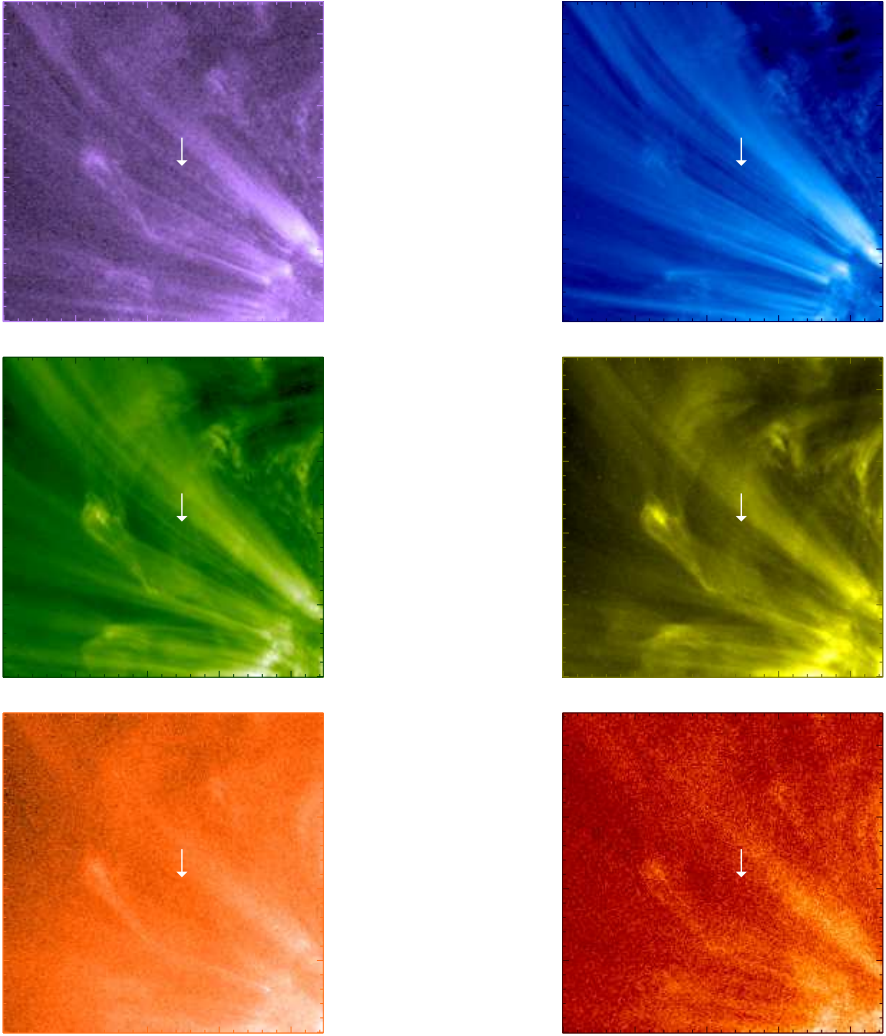}
\caption{Rainbow Plot with a format similar to Figure 2 showing the subsection of AR 11944 containing Region E Loop 9. Loop 9 is categorized as effectively isothermal.
}
\end{figure}

\clearpage

\begin{figure}
\epsscale{.7}
\plotone{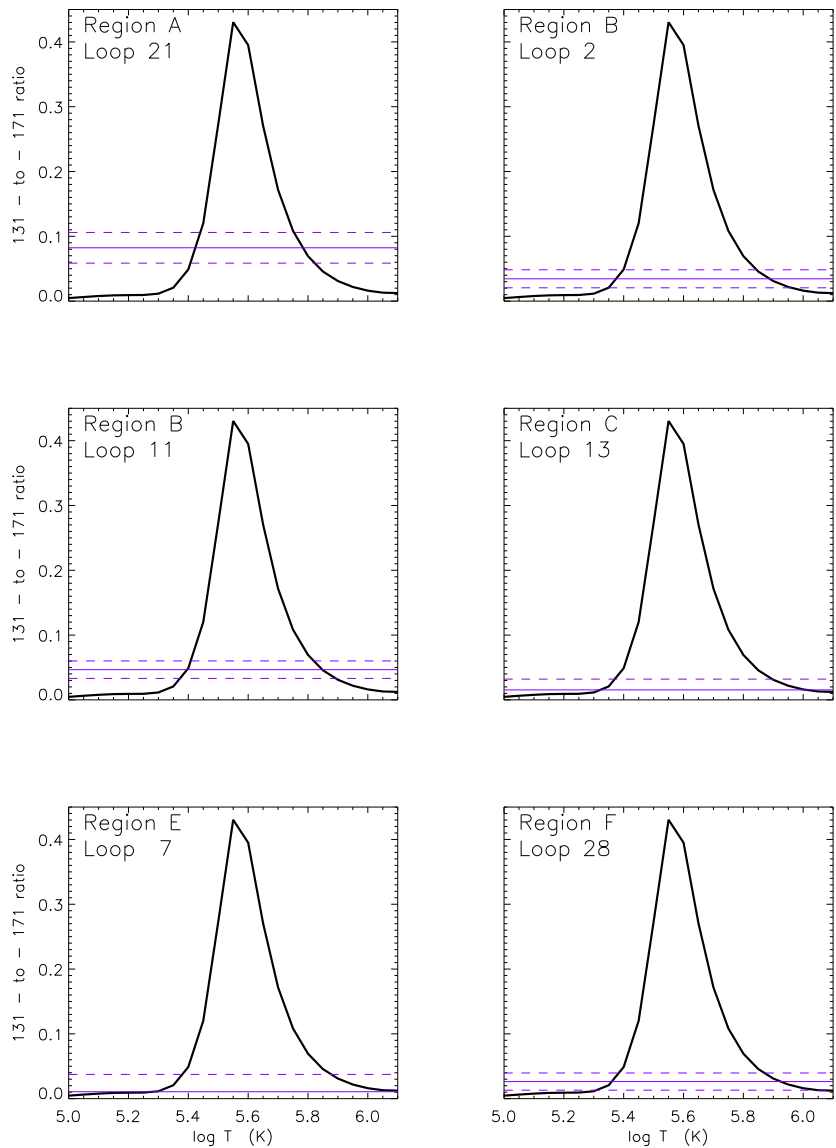}
\caption{For each panel, the 131-to 171 AIA instrument response ratio is in black with a peak at Log T = 5.55. The observations are in purple. The solid flat lines show the 131-to-171 average background-subtracted intensity ratio and the dashed lines show the uncertainties. Because each horizontal line intersects twice with the response ratio, there are two possible temperatures for each loop segment that are consistent with the observations, one at Log T $\sim$ 5.35 and the other at Log T $\sim$ 5.95.
}
\end{figure}

\clearpage

\begin{figure}
\epsscale{.9}
\plotone{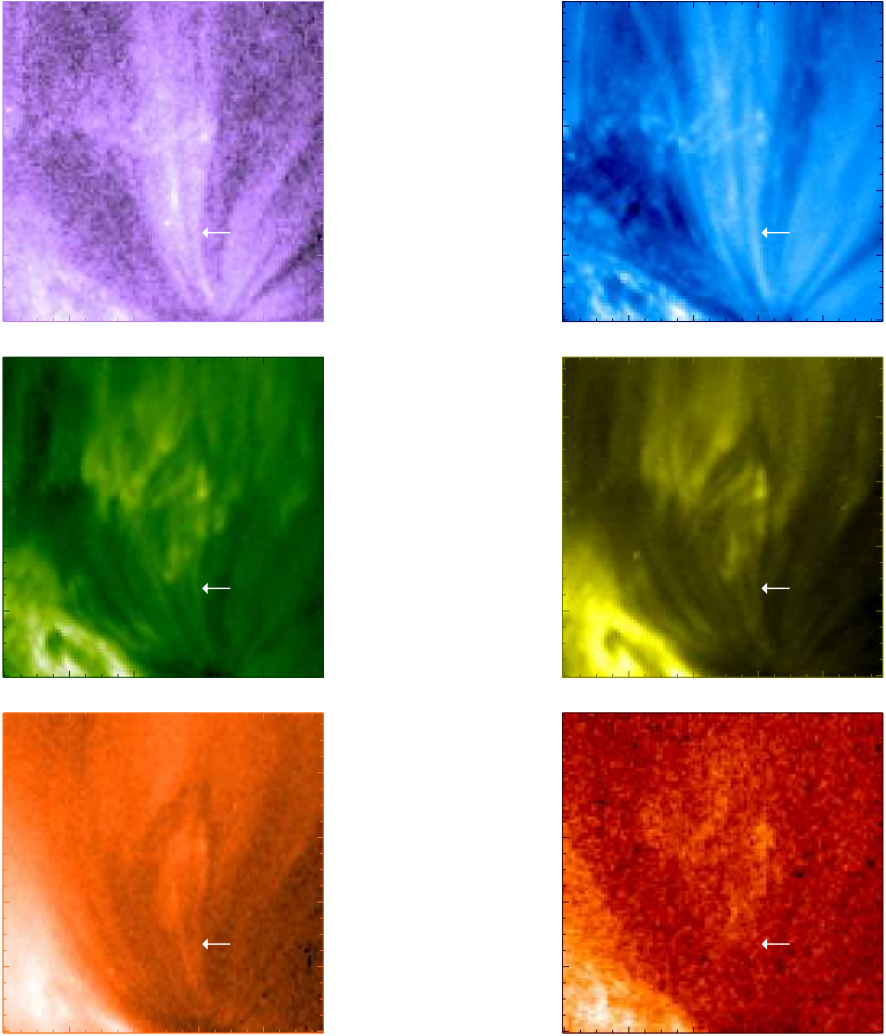}
\caption{Rainbow Plot similar to Figure 2 showing the subsection of Rainbow Plot showing the subsection of AR 11166 containing Region B Loop 43. Loop 43 is an example of a multithermal loop.
}
\end{figure}

\clearpage

\begin{figure}
\epsscale{.9}
\plotone{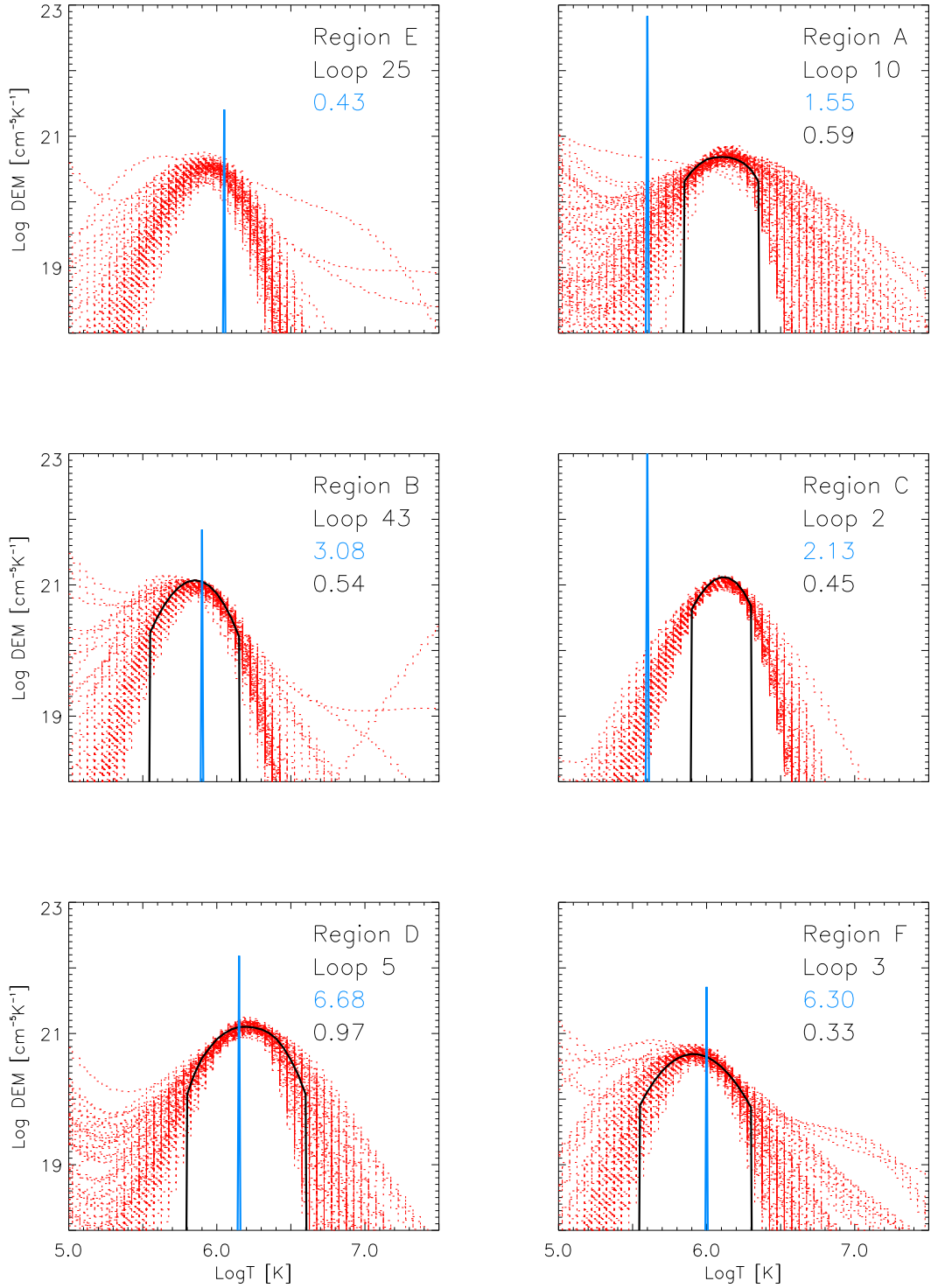}
\caption{Results from DEM\_manual for the best isothermal (blue) and multithermal (black) fit to the background-subtracted loop data. The reduced ${\chi^2}$ for each model is listed in the upper right corner of each panel using the same color coding. The Monte Carlos from xrt\_dem\_iterative2 (red) are overplotted.
}
\end{figure}

\clearpage

\begin{figure}
\epsscale{.9}
\plotone{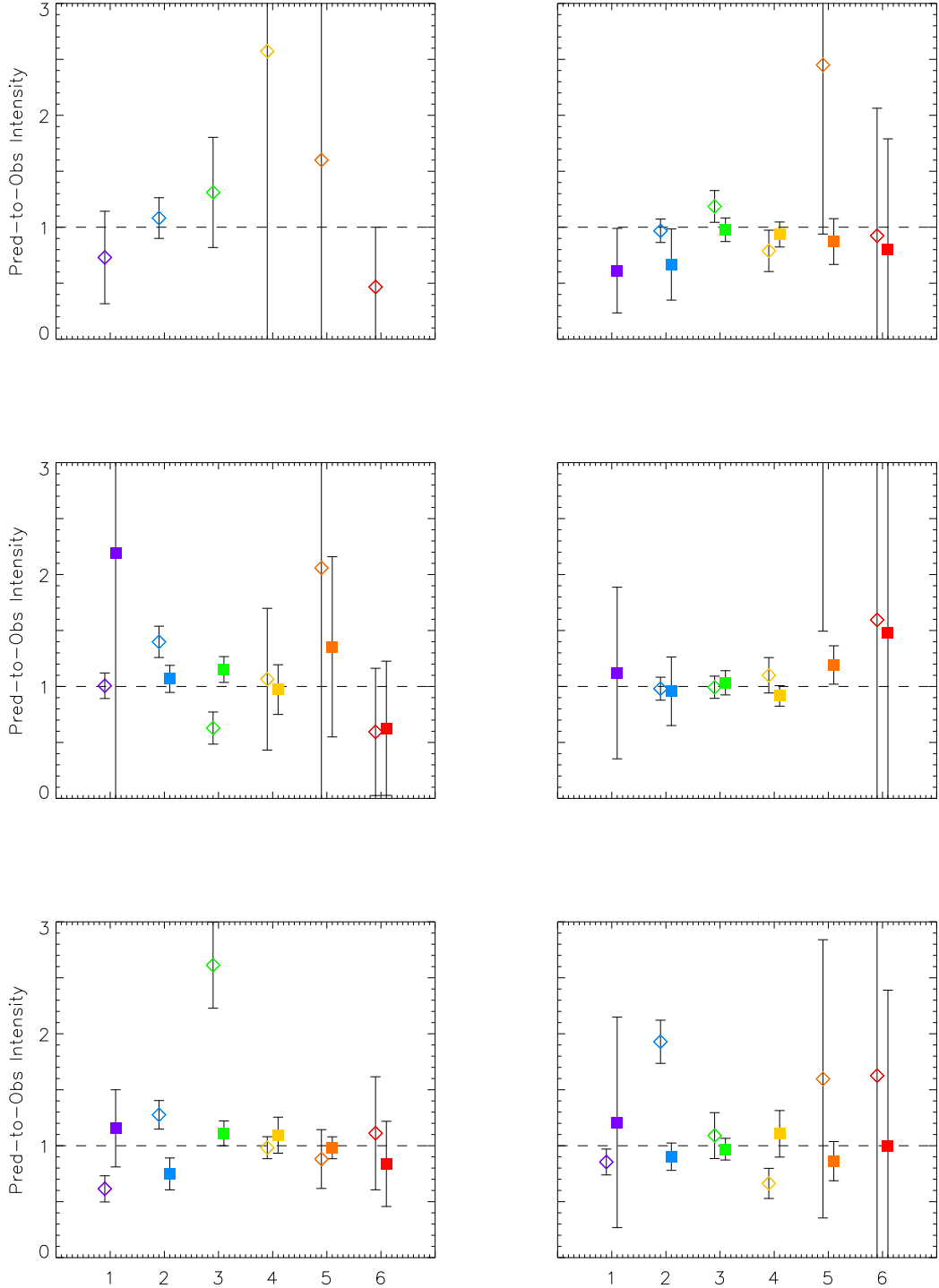}
\caption{Predicted-to-observed intensity ratios for the isothermal (diamonds) and multithermal (squares) DEM models from Figure 7. The data points show the six AIA coronal filters in approximate temperature order: (1) 131 \AA, (2) 171 \AA, (3) 193 \AA, (4) 211 \AA, (5) 335 \AA, (6) 94 \AA. The color scheme is the same as that used in Figure 2. Note: the large error bars result when the loop intensity is barely above background.
}
\end{figure}

\end{document}